# Analyses of Microstructural and Elastic Properties of Porous SOFC Cathodes Based on Focused Ion Beam Tomography

Zhangwei Chen*, Xin Wang, Finn Giuliani and Alan Atkinson
Department of Materials, Imperial College London SW7 2BP, UK

[*] Corresponding author, currently in Department of Earth Science and Engineering, Tel.: +44 2075949695; Fax: +44 2075949625; E-mail address: chen@ic.ac.uk (Z. Chen)

**Abstract**

Mechanical properties of porous SOFC electrodes are largely determined by their microstructures. Measurements of the elastic properties and microstructural parameters can be achieved by modelling of the digitally reconstructed 3D volumes based on the real electrode microstructures. However, the reliability of such measurements is greatly dependent on the processing of raw images acquired for reconstruction. In this work, the actual microstructures of $La_{0.6}Sr_{0.4}Co_{0.2}Fe_{0.8}O_{3-\delta}$ (LSCF) cathodes sintered at an elevated temperature were reconstructed based on dual-beam FIB/SEM tomography. Key microstructural and elastic parameters were estimated and correlated. Analyses of their sensitivity to the grayscale threshold value applied in the image segmentation were performed. The important microstructural parameters included porosity, tortuosity, specific surface area, particle and pore size distributions, and inter-particle neck size distribution, which may have varying extent of effect on the elastic properties simulated from the microstructures using FEM. Results showed that different threshold value range would result in different degree of sensitivity for a specific parameter. The estimated porosity and tortuosity were more sensitive than surface area to volume ratio. Pore and neck size were found to be less sensitive than particle size. Results also showed that the modulus was essentially sensitive to the porosity which was largely controlled by the threshold value.

## 1. Introduction

Solid oxide fuel cells (SOFCs) are promising energy conversion devices which directly generate electricity from the electrochemical reactions of fuels and air (oxygen) with high efficiency and low environmental impact [1]. A SOFC commonly consists of a dense ceramic electrolyte with high oxygen ionic conductivity, which is supported on either side by a porous cathode with mixed ionic-electronic conductivity (MIEC) and a porous anode, respectively. Perovskite-structure materials such as $La_{0.6}Sr_{0.4}Co_{0.2}Fe_{0.8}O_{3-\delta}$ (LSCF) have been widely applied as SOFC cathodes due to their good MIEC, particularly at intermediate operating temperatures [2-5].

The electrochemical performance and mechanical properties of an operating LSCF cathode are strongly dependent on their three-dimensional (3D) microstructures [6, 7], which include the porosity, distributions of particle and pore sizes, surface area and phase tortuosity. Often larger porosities are favourable for enhancing the cathode's electrochemical performance while its mechanical strength increases when the opposite is the case. Therefore, efforts must be made to trade off the electrochemical performance and the ability to withstand mechanical constraints.

The development of advanced 3D tomographic techniques has made it possible to analyse real spatial microstructures of SOFC electrodes by 3D reconstruction [8]. The reconstructed actual 3D microstructures allow mechanical and electrochemical simulations which can help improve the mechanical and electrochemical properties. One of the most widely used techniques is the focused ion beam/scanning electron microscope (FIB/SEM) tomography, which was first applied to the characterisation of SOFC materials by Wilson et al. [9] and has been implemented to reconstruct 3D microstructures of SOFC anodes thanks to its advanced slicing and imaging capability, and high spatial resolution. With the aid of the FIB/SEM technique, studies have been reported of the 3D reconstruction of LSCF cathode microstructures and the relationship to their electrochemical performance [10-12]. However, none of these 3D reconstructions of porous thin LSCF cathodes was correlated to their mechanical properties, which on the other hand has been rarely studied [7], as most mechanical studies in the literature were based on nominally dense LSCF bulk samples [13-15]. However, porous films may behave mechanically very differently from bulk samples of the same materials/compositions.

On the other hand, reliable quantification of the microstructural parameters and mechanical/electrochemical simulations require accurate 3D microstructure datasets, which hinge largely on the quality of the binarised 2D sequential images of the sample volume. The segmentation process, which generally involves grayscale thresholding to identify the two phases (i.e. pore and solid phase), is crucial to the resulting binary image quality and thus the 3D microstructures reconstructed.

The current study aims to investigate the sensitivity of key microstructural and elastic parameters to the grayscale threshold value applied in the image segmentation process. The important microstructural parameters involved porosity, tortuosity, specific surface area, particle and pore size distributions, and inter-particle neck size distribution, all of which may, to different extents, influence the elastic properties (i.e. elastic modulus and Poisson's ratio) calculated by finite element modelling based on the reconstructed 3D microstructures. The input Poisson's ratio of dense LSCF material was also examined to check its influence on the elastic parameters of the porous microstructures.

## 2. Materials and Methods

### 2.1. Sample Preparation

LSCF cathode films were fabricated by tape casting of an ink slurry on CGO ($Ce_{0.9}Gd_{0.1}O_{2-\delta}$) pellet substrates, followed by sintering at 900 °C in air for 4 hours. The resulting films had smooth and crack-free surface, without interfacial delamination from the substrate. The detailed sample fabrication processes can be found in [7, 16].

### 2.2. FIB/SEM Slicing and Viewing for Image Acquisition

The as-sintered porous films were coated with a thin gold layer on the top and subjected to slicing and viewing with a combined focused ion beam and scanning electron microscope, i.e. FIB/SEM instrument (Helios NanoLab 600, FEI, USA), for acquiring sequential cross-sectional 2D high definition SEM images of the films. Prior to the gold coating, the films were vacuum impregnated with epoxy resin such that the highly interconnected porous structures close to the slicing area could be retained during slicing and at the same time an improved contrast could be generated for better distinction between pore and solid phases.

The resolution of the image stacks in the normal direction of the image plane was equivalent to the predefined slice thickness, which ideally should be set to be the same as the resolution of the image, i.e. pixel size. This was done so that cubic voxels could be generated to facilitate the ensuing reconstruction and simulation. Otherwise, the resolution can be rectified by downsampling of the FIB/SEM data, which would however lose some information of the original images. In the present study, the slice thickness and the image pixel size were all set to be 12.5 nm, which resulted in a voxel size of 12.5 nm cube side after reconstruction. The typical beam voltage and current used in this study are listed in Table 1. Note that here "Patterning" refers to rough and quick milling of large trenches around the volumes of interest (VOIs), while "Deposition" of thin layers of platinum or carbon on the top of the VOIs helped reduce the so-called curtain effect which resulted in poor cross-sectional surface topography. The remaining working parameters were: FIB working distance = 4 mm; imaging scanning speed (i.e. imaging time per spot of test material) = 300 ns to 10 μs and tilting angle = 52°.

*Table 1 Typical beam voltage and current used of the FIB/SEM in this study.*

| Beam Type | Voltage | Current |
|---|---|---|
| SEM | 5-15 kV | Imaging: 0.17 nA |
| FIB | 30 kV | Imaging: 93 pA<br>Patterning: 9.3-21 nA<br>Deposition: 0.92 nA<br>Slicing: 0.92 nA |

### 2.3. 3D Microstructure Reconstruction

The actual 3D volumes of the films were then reconstructed based on the as-recorded stacks of images using Avizo 8.0 software (VSG, FEI, USA). The reconstruction involved processing of the images to achieve image alignment and binarisation of pore and solid phases (i.e. segmentation). The segmentation was done by grayscale thresholding of the as-acquired grayscale images into binary images, where pixels with grayscale values larger than the pre-chosen threshold value (denoted as *TV*) were labelled as the foreground (solid phase) and pixels with grayscale values smaller than the *TV* were classified as background (pore phase). The sensitivity of these parameters to the *TV* used was investigated by varying the *TV* within a limited range close to the minimum value between the two peaks of the corresponding grayscale histogram of the images. This will be discussed and analysed in detail later.

### 2.4. Quantification of Microstructural Parameters

Key microstructural parameters including porosity, tortuosity, surface area to volume ratio (denoted as *SAVR*), and particle and pore size distributions, and inter-particle neck size distribution were measured using Avizo's quantification module. The porosity was calculated simply as the fraction of voxel number possessed by pore phase over the total voxel number of the volume. In a 3D context tortuosity is a parameter describing the extent of a twisted interconnected path throughout the volume. In the current study, tortuosity of both pore and solid phases was measured using geometrical approach, based on a computed path formed by centroids of each interconnected region identified as same phase on each image of the dataset along the FIB slicing direction. As a result, tortuosity was calculated by dividing the path length through the centroids by the number of images along the slicing direction which represented the length between the two ends of the path. The surface to volume ratio of the microstructures was determined as the ratio of the pore (or solid) phase surface area to the total volume. The quantification of particle and pore sizes were achieved by applying an object separation operation to the microstructures with a watershed algorithm [17], which partitioned interconnected solid (or pore) phase into a group of individual particles (or pores). Here, the volume of each particle (or pore) was measured and converted into an *Equivalent Spherical Diameter (ESD)* for plotting the particle (or pore) size distribution histogram. *ESD* of a particle/pore in 3D can be obtained by $ESD = (6V/\pi)^{1/3}$, where *V* denotes the particle/pore volume. The sizes of inter-particle necks, which were 2D interfaces between any two originally interconnected particles, were also measured after partitioning was performed. Here, the surface area of each individual neck was measured and converted into an *Equivalent Circular Diameter (ECD)* for neck size distribution analysis. The *ECD* of an interfacial neck can be derived by $ECD = 2(A/\pi)^{1/2}$, where *A* denotes the neck surface area.

### 2.5. Finite Element Modelling

Elastic modulus and Poisson's ratio are two fundamental parameters to characterise the elastic properties of materials and thus are essential to know. The former indicates the resistance of a material to elastic deformation upon application of force and it is defined as the ratio of stress to strain in the elastic deformation region. The negative ratio of a material's transverse strain to axial strain is referred to as Poisson's ratio and this constant, ranging often between 0.0 and 0.5, reflects transverse deformation caused by axial stress. 3D finite element models were generated by applying adaptive tetrahedral meshing to the as-reconstructed 3D microstructures using ScanIP (Simpleware, UK). Mechanical simulations for calculating the effective elastic modulus and the Poisson's ratio of each individual microstructure were carried out using Abaqus CAE 6.10 (Dassault Systemes, USA) in a standard analysis mode, on the assumption that LSCF solid material was isotropic, linear and elastic. In the simulations, a Poisson's ratio of fully dense LSCF material reported in literature [13], $v_s$ = 0.30, was chosen as the base case for use in FEM simulations. An elastic modulus of fully dense LSCF material determined by nanoindentation in previous study [7] was chosen as the solid phase elastic parameter: $E_0$ = 175 GPa. A small displacement was applied on one free surface normal to the $Y$ axis (normal to the film plane, as indicated in Fig. 1) so that the model deformed linearly along $Y$. The opposite face was constrained to have no displacement in this direction. Boundary conditions were also applied to constrain the degree of freedom of the normal displacement for the nodes on the model's other surfaces parallel to $Y$. Such settings allowed these surfaces to contract or extend freely once the displacement was applied. The resultant normal force on the displaced $Y$ surface was obtained from the model so that the effective elastic modulus of the 3D microstructure could be determined by simply dividing the resultant force by the total area (solid plus pore) $L_x \times L_z$ of the displaced surface. The Poisson's ratio could be deduced as well by calculating the absolute ratio of the corresponding transverse strain over the axial strain. The sensitivity of the effective elastic modulus and the Poisson's ratio of the porous microstructures to the input solid Poisson's ratio was also tested.

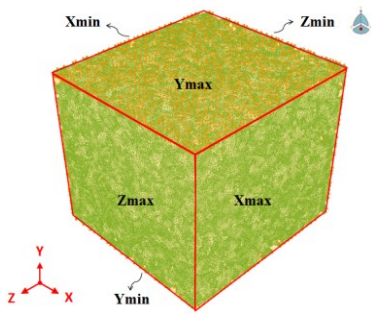

*Fig. 1 A meshed model under application of boundary conditions.*

Fig. 2 summarises the workflow of the 3D microstructure reconstruction, mesh generation and FEM simulation processes including applications and software used, the resulting output file types as well as the file extensions.

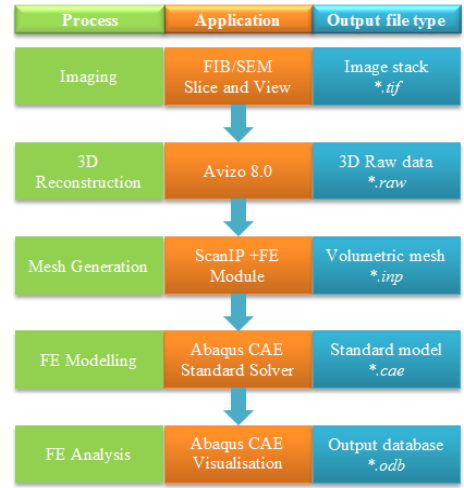

*Fig. 2 Workflow of dataset processing for image analysis and finite element modelling.*

### 3. Results and Discussion
#### 3.1. 3D Microstructure Reconstruction

In the current study, three volumes of interest were sampled in different locations of the films sintered at 900 ˚C. Fig. 3 (a) shows an image stack consisting of 250 slices obtained by FIB/SEM slicing and viewing with image size of 250×250 pixel$^2$. Fig. 3 (b) is the corresponding histogram which plots the number of voxels of each grayscale value for the above image stack, with 0 corresponding to black (i.e. pore phase in Fig. 3 (a)). It shows a reasonably good bimodal feature which was advantageous for subsequent image segmentation. Thus the minimum value (=36) between the two peaks was chosen to be the initial *TV* for segmentation. Segmentation using the Otsu algorithm [18] performed by the ImageJ plugin "Otsu_Threshoding.jar" [19] resulted in a *TV* = 35, almost identical to the one obtained above, suggesting the minimum value a proper *TV*.

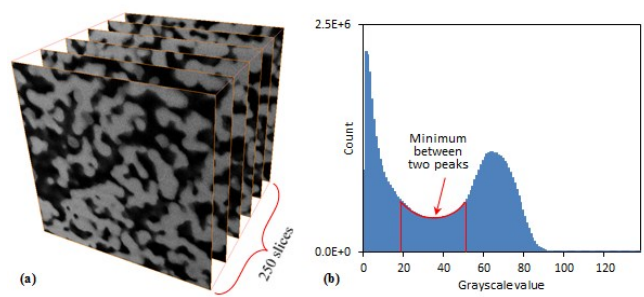

*Fig. 3 Example of a microstructure image stack of a specimen sintered at 900 ˚C (a), and the corresponding grayscale histogram (b). Note that the minimum between the two peaks is at the grayscale value 36 which was taken as the initial TV for image segmentation.*

In order to evaluate the sensitivity of the key microstructural parameters to the variation of *TV*, a number of grayscale values within a limited range close to 36 were chosen to be used for segmentation, i.e. 19, 25, 31, 36, 43, 49 and 55. The example of the results after applying the above

*TVs* is illustrated in Fig. 4, which shows a grayscale image from the stack and the resulting binary images generated when different *TVs* were applied for segmentation. The corresponding 3D microstructures reconstructed are shown in Fig. 5.

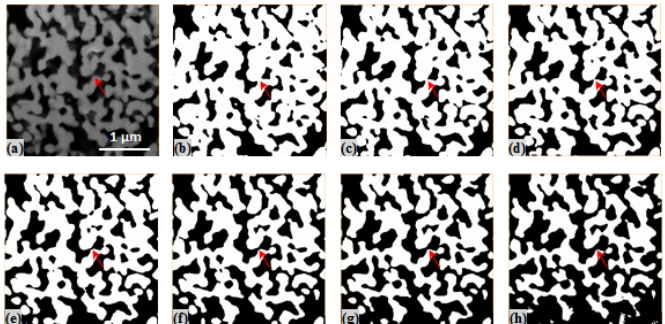

*Fig. 4 Example of a grayscale image and the corresponding segmented binary images at different TVs: (a) initial grayscale image, and binary image by applying a TV of (b) 19, (c) 25, (d) 31, (e) 36, (f) 43, (g) 49 and (h) 55. An ambiguously non-connected solid feature is arrowed in each image to show the difference was made by applying different TVs.*

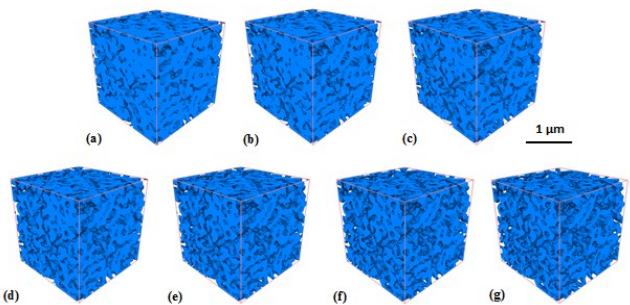

*Fig. 5 The corresponding reconstructed 3D volumes of the segmented image stack by applying a TV of (a) 19, (b) 25, (c) 31, (d) 36, (e) 43, (f) 49 and (g) 55.*

It can be seen from the images above that as the *TV* was chosen to vary from 19 to 55, an increasing number of voxels were identified as background pore phase, which would result in increased porosity in the microstructures, as also can be readily seen in the 3D microstructures shown in Fig. 5. Note that the relatively non-uniform binary image shown in Fig. 4 (h) might be due to a relatively minor contrast gradient across the field view during image acquisition. A typical feature of the solid phase that is sensitive to the *TV* is arrowed in Fig. 4 which shows how apparent phase connectivity can be modified by changes in *TV*.

### 3.2. Sensitivity of Microstructural Parameters to TV
#### 3.2.1. Porosity and SAVR

Key microstructural parameters such as porosity, *SAVR*, tortuosity, as well as microstructural feature size including particle and pore size distributions and inter-particle neck size distribution during the course of the *TV* variation were all calculated and plotted for the three volumes of interest. The measurements of each individual parameter in the three volumes all show very consistent results, as described below.

Fig. 6 shows the changes of porosity and *SAVR* as the applied *TV* varied from 19 to 55. As suggested by the readily seen change of the segmented microstructure shown in Fig. 4 and Fig. 5, Fig. 6 confirmed that the apparent porosity increased significantly with the increase of the *TV*, consistent with the results reported by Joos et al. [11] working on the similar material. Indeed, an increasing number of voxel were considered to belong to the pore phase when higher *TVs* were used for image segmentation. The apparent porosity almost doubled from approximately 30 % to 60 % when *TVs* of 19 and 55 were used, respectively, showing a great sensitivity of porosity to the *TV*. On the other hand, the plots for *SAVR* show little influence of the *TV* variation. The averaged porosity and *SAVR* for the three volumes of interest at *TV* = 36 were measured to be 45.2 ± 2.0 % and 6.9 ± 0.4 μm$^{-1}$, respectively. These are very close to the measurements reported by Joos et al. [11] for their LSCF cathode, with porosity of approximately 48 % and *SAVR* of 6.2 μm$^{-1}$ at a proper *TV* chosen.

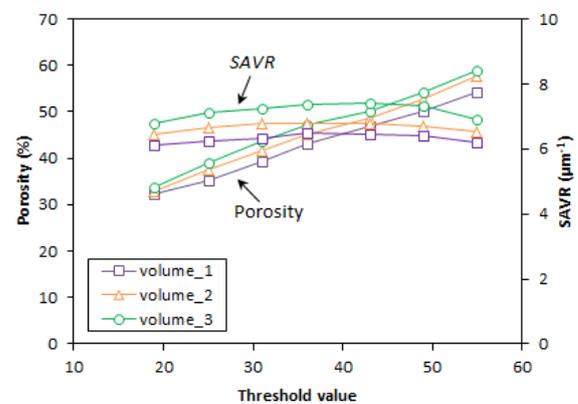

*Fig. 6 Variations of porosity and SAVR for the reconstructed 3D volumes with different TV ranging from 19 to 55. Note that the minimum TV is 36.*

In order to gain further insight into the sensitivities of the two parameters to the *TV*, the changes of porosity and *SAVR* were also plotted relative to the *TVs* over the whole range of grayscale value present in the histogram for one of the three volumes, namely from 0 to 115, as shown in Fig. 7.

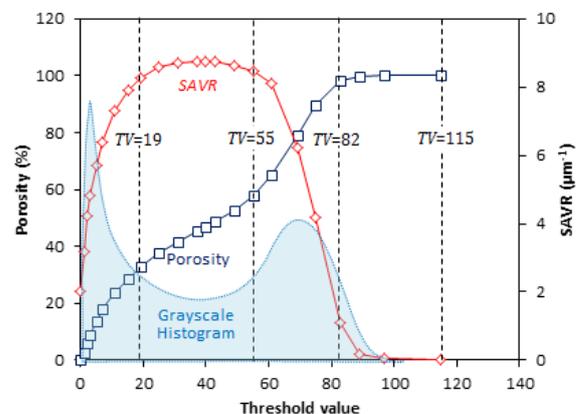

*Fig. 7 Variations of porosity and SAVR for a reconstructed 3D volume with different TV ranging from 0 to 115.*

Fig. 7 reveals that the sensitivity is strongly dependent on the range of threshold values chosen. It can be noticed that the most sensitive *TV* intervals shown in the figure for both parameters are $0 \leq TV \leq 19$ and $55 \leq TV \leq 82$ coinciding with the peaks in the grayscale histogram. For $82 \leq TV \leq 115$, the whole microstructure became nominally void without any solid phase present, with a constant porosity of 100 % and hence obviously the *SAVR* became zero. According to the analysis and discussion above, it can be concluded that the two parameters were extremely sensitive to the *TVs* in the peak regions of the grayscale histogram. In the current study, only the interval $19 \leq TV \leq 55$ was considered for analysis of other parameters that followed.

*3.2.2. Tortuosity*

It should be noted that for a volume of interest, the tortuosities of solid and pore phases may not be the same. Tortuosity cannot be less than 1 and microstructures with more twisted and complex features result in higher tortuosity values. In the current study both pore and solid tortuosities were measured and the results are plotted in Fig. 8.

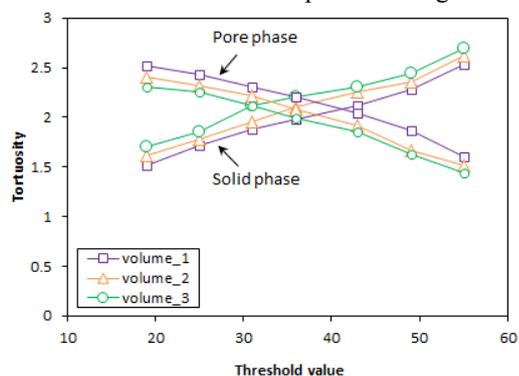

*Fig. 8 Variations of apparent tortuosities for both pore and solid phase as a function of TV.*

The figure shows an increase of solid tortuosity from 1.5 to approximately 2.5 and a corresponding decrease of apparent pore tortuosity to 1.5 from 2.5, as the *TV* increased. As the *TV* increased, more and more voxels in the microstructures were classified as the pore phase, which became increasingly interconnected and less twisted. As a result, the tortuosity decreased. While for the solid phase, the opposite was true. The averaged tortuosities for both pore and solid phases were measured to be the same at *TV*=36, with a value of 2.1 ± 0.1. Such a similarity in tortuosities is not surprising, because the porosity at *TV*=36 was approximately half of the volume (i.e. 45.2 ± 2.0 %, as shown before) and similar geometrical characteristics were expected to be present for both pore and solid phases in the resulting microstructures at this *TV*, as can be seen in Fig. 4 and Fig. 5.

Tortuosity also links the bulk conductivity to the conductivity of a specific phase structure. Therefore, it can also be determined based on transport equation describing the simulation of fluid diffusion in the entire media of the structure phase [20]. The tortuosities calculated in the current study remained fairly close to the data reported by Joos et al. [11], which are 2.05 and 1.94 for LSCF material and pores, respectively at the properly chosen *TV*, using a homogenisation approach-based transport simulation.

*3.2.3. Particle Size, Pore Size and Inter-particle Neck Size Distributions*

Besides porosity, features such as sizes of particles, pores and the inter-particle necks are crucial for investigating their underlying correlation with elastic properties of the microstructures [21-25]. The sizes of these important features were calculated and plotted as frequency distributions for different *TVs* used, as shown in Fig. 9, Fig. 10 and Fig. 11. Fig. 12 plots the averaged values (with standard deviations) calculated from the size distributions shown in these figures.

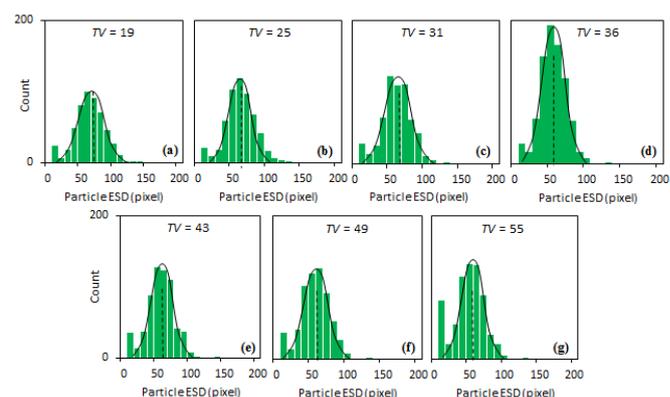

*Fig. 9 Particle size (ESD) distribution at different threshold values applied (1 pixel = 12.5 nm): (a) TV=19, (b) TV=25, (c) TV=31, (d) TV=36, (e) TV=49, (f) TV=43 and (g) TV=55.*

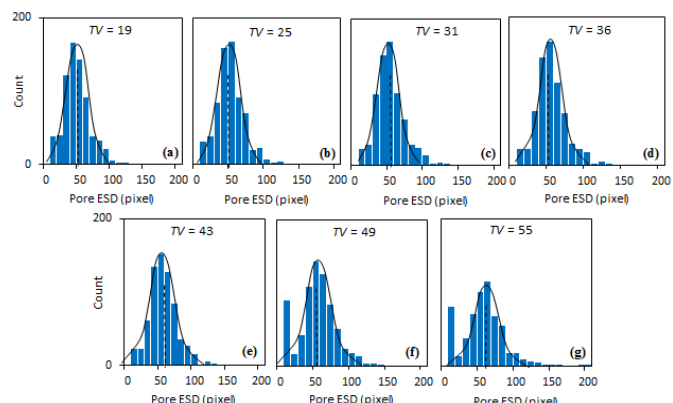

*Fig. 10 Pore size (ESD) distribution at different threshold values applied (1 pixel = 12.5 nm): (a) TV=19, (b) TV=25, (c) TV=31, (d) TV=36, (e) TV=49, (f) TV=43 and (g) TV=55.*

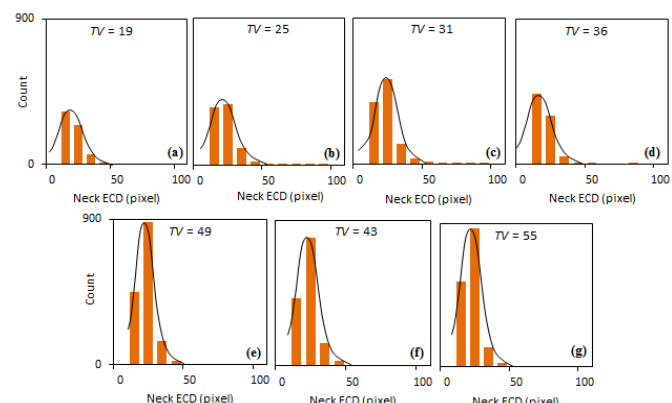

*Fig. 11 Inter-particle neck size (ECD) distribution at different threshold values applied (1 pixel = 12.5 nm): (a)*

TV=19, (b) TV=25, (c) TV=31, (d) TV=36, (e) TV=49, (f) TV=43 and (g) TV=55.

It can be seen from Fig. 9 and Fig. 10 that the size distributions of both particles and pores are quite similar, with unimodal feature being mostly present. However, particles exhibited broader size distribution when lower *TVs* were used, whereas pores had wider size distribution when larger *TVs* were used. This is also indicated by the standard deviations of the corresponding data plots shown in Fig. 12.

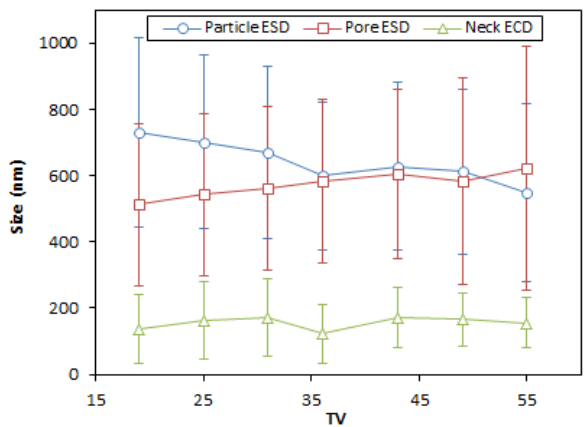

*Fig. 12 Average apparent particle, pore and neck sizes as a function of the TV used.*

Generally particles tended to have larger mean apparent size than pores, particularly when lower *TVs* were used. They became closer for *TV* above 36. The mean apparent size of particles showed a decrease over the *TV* range studied by approximately 25%, whereas at the same time the mean apparent size of pores rose by 21%. However, the mean apparent neck size was almost constant regardless of the variation of *TV*. It is worth noticing that the large standard deviations were not caused by the negligible difference of the three sample volumes, but mainly due to the extremely non-uniform sizes of particles/pores/necks in each of the volumes. Although the averaged values can be used in an indicative way for comparison, it is worth noticing that the large standard deviation shown in Fig. 12 suggests that it is more appropriate to represent the feature sizes as frequency distributions as shown in Fig. 9, Fig. 10 and Fig. 11 rather than looking at the mean particle size only.

*3.3. Sensitivity of Effective Elastic Modulus and Poisson's Ratio to TV and Input Poisson's Ratio*
*3.3.1. Sensitivity to the Input Poisson's Ratio*

Fig. 13 (a) and (b) shows respectively the effective elastic moduli and Poisson's ratios of the as-reconstructed microstructure calculated using FEM at *TV*=36 as a function of the input Poisson's ratio defined for the solid LSCF phase, i.e. $v_{LSCF}$, which varied from 0.15 to 0.45.

It is readily seen from the figures above that the increase of input $v_{LSCF}$ led to increase in both effective elastic parameters of the microstructure. However, for the effective elastic modulus, merely less than 10% of increase was experienced with the increase of $v_{LSCF}$ from 0.15 to 0.45. On the other hand, in terms of the resulting effective Poisson's ratio, fairly linear increase can be observed from 0.14 to 0.2 as $v_{LSCF}$ rose from 0.15 to 0.45. Therefore compared with the resulting effective Poisson's ratio, the effective elastic modulus was less sensitive to the input $v_{LSCF}$.

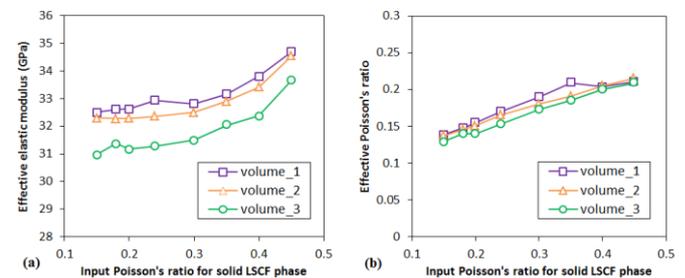

*Fig. 13 Effective elastic moduli (a) and Poisson's ratios (b) of the as-reconstructed microstructure calculated by using different input Poisson's ratios for solid LSCF phase.*

Mechanical properties and electrochemical performance are two main aspects that need to be considered when designing an electrode component in SOFCs. They are greatly dependent on the microstructural characteristics of the porous component and are generally contradicting each other. Because a relatively larger porosity is usually more favourable for enhanced electrochemical performance, whereas, to maintain durable mechanical properties a lower porosity is needed. In FEM simulations, the input Poisson's ratio for a solid material is not always known. An unreasonable blind guess would often lead to uncertainty of the calculation. The results here suggest that the use of $v_{LSCF}$ in the range close to 0.30 could be reasonable for the calculation to be reliable, in case that the $v_{LSCF}$ might be ambiguously reported in the literature. Reliable FEM-derived mechanical properties which are comparable to the experimental mechanical testing could help to effectively balance the trade-off of microstructural parameters by taking into account both mechanical and electrochemical performance.

*3.3.2. Sensitivity to TV*

The effective elastic modulus and the Poisson's ratio for each microstructure were also computed with the input Poisson's ratio being 0.3 over the variation of *TV* ranging from 19 to 55 and results are plotted in Fig. 14.

As can be seen from Fig. 14, the effective elastic modulus experienced a significant drop from approximately 60 GPa to 15 GPa, with the increase of *TV* used from 19 to 55, due to the increasing amount of apparent porosity. While for the Poisson's ratio, only a minor reduction took place over the *TV* range, resulting in the Poisson's ratio of approximately 0.24 at *TV*=19 and 0.23 at *TV*=55. Indeed, the effective elastic modulus was essentially sensitive to the porosity which was largely dependent on the *TV* used. However, the increase of porosity did not necessarily have strong influence on the variation of the Poisson's ratio. The effective elastic modulus and the Poisson's ratio at *TV*=36 were measured to be 32.3 ± 0.7 GPa and 0.24 ± 0.01, respectively. Such an estimated elastic modulus has been proven consistent with experimental data measured by a nanoindentation technique, as reported by our previous work [7].

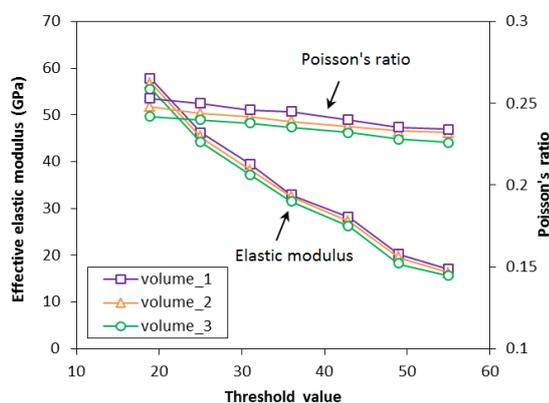

*Fig. 14 Variations of effective elastic modulus and Poisson's ratio as a function of TV.*

More details regarding the resulting 3D models and the relating FE mechanical simulation are given in Table 2 as a function of *TV* used. It can be seen that as the *TV* increased, the apparent volume of solid content (i.e. LSCF) reduced, so that the numbers of solid voxels, meshed elements and nodes also decreased. This resulted in increasingly smaller dataset size as well as shorter time to complete the simulation (which was run by a workstation configured with Intel Xeon 12-core 2.76 GHz processor and 96 GB RAM).

*Table 2 Detailed modelling information of the 3D models as the TV used varied from 19 to 55.*

| TV | LSCF Voxel Number ($\times 10^6$) | Element Number ($\times 10^6$) | Node Number ($\times 10^5$) | File Size (MB) | Simulation Time (mins) |
|---|---|---|---|---|---|
| 19 | 10.50 | 2.98 | 6.41 | 148.1 | 8.8 |
| 25 | 9.75 | 2.53 | 5.86 | 130.9 | 8.3 |
| 31 | 9.13 | 2.31 | 5.58 | 121.2 | 7.2 |
| 36 | 8.87 | 2.01 | 5.13 | 107.1 | 6.3 |
| 43 | 8.00 | 1.88 | 5.01 | 99.70 | 5.5 |
| 49 | 7.38 | 1.74 | 4.75 | 92.90 | 5.2 |
| 55 | 6.61 | 1.65 | 4.58 | 88.20 | 5.0 |

## 4. Conclusions

The mechanical properties of porous thin SOFC electrode films strongly rely on some of their key microstructural parameters, which can be made available by quantification based on 3D reconstructed microstructures using FIB/SEM tomography. Reliable microstructural parameter quantification and mechanical simulation require accurate processing of the as-collected sequential 2D image stacks. The segmentation process, which generally involves grayscale thresholding to identify the two phases (i.e. pore and solid phase), is one of the most crucial steps to affect the resulting binary image quality and thus the 3D microstructures reconstructed.

In this work, the actual microstructures of $La_{0.6}Sr_{0.4}Co_{0.2}Fe_{0.8}O_{3-\delta}$ (LSCF) cathodes sintered at an elevated temperature were reconstructed, resulting in models with a very large number of voxels (~ 10 million). The sensitivity of key microstructural (i.e. porosity, tortuosity, specific surface area, and particle, pore and neck size distributions) and elastic parameters (i.e. elastic modulus and the Poisson's ratio) to the grayscale threshold value applied in the image segmentation process was studied. The elastic parameters were computed using finite element modelling based on the actual reconstructed 3D microstructures. Results have shown that different *TV* range would result in different degree of sensitivity for a specific parameter.

The *TV* value was varied in the valley region of the histogram where an effective and reasonable *TV* was located (in this study $19 \leq TV \leq 55$). The porosity and tortuosity of the phases were more sensitive than surface area to volume ratio. The particle size was more sensitive than the pore and neck size. The apparent particle mean size decreased by approximately 25%, whereas the apparent pore mean size rose by 21%. However, the neck size was relatively insensitive to the *TV*. The effective elastic modulus was more sensitive than the Poisson's ratio. The modulus was essentially sensitive to the apparent porosity which was largely controlled by the *TV* used and dropped from 60 to 15 GPa, as the *TV* increased from 19 to 55. However, the apparent Poisson's ratio almost kept constant at 0.24.

The sensitivity of the computed elastic parameters to the variation of input Poisson's ratio of the solid LSCF $v_{LSCF}$ was also studied. Results showed that compared with the effective Poisson's ratio, the effective elastic modulus was less sensitive to the input $v_{LSCF}$.

In the current study, a more realistic range of *TVs* could be that with which agreement can be reached when comparing elastic moduli measured using both 3D microstructure-based FEM and experimental nanoindentation method. Therefore, if a variation of ± 20 % of the "true" elastic modulus is acceptable (i.e. elastic modulus = 32 ± 6 GPa), then it can be found in Fig. 14 that the lower and upper limits of *TVs* are approximately 30 and 40, assuming that the elastic modulus to *TV* relationship is linear in this range.

As a result, the microstructural parameters and effective elastic parameters at *TV*=36 were measured to be porosity 45%; surface area to volume ratio 6.9μm$^{-1}$; tortuosity for both phases 0.21; effective elastic modulus 32 GPa; and Poisson's ratio 0.24, agreed with nanoindentation experimental results.

**Acknowledgements**

This research was carried out as part of the UK Supergen consortium project on "Fuel Cells: Powering a Greener Future". The Energy Programme is an RCUK cross-council initiative led by EPSRC and contributed to by ESRC, NERC, BBSRC and STFC. Z. Chen is especially grateful to the Chinese Government and Imperial College for financial support in the form of scholarships.